\documentclass[a4paper,12pt]{article}

\usepackage[a4paper,margin=2.5cm]{geometry}
\usepackage[utf8]{inputenc}
\usepackage{mathtools}
\usepackage{amsmath}
\usepackage{braket}
\numberwithin{equation}{section}
\usepackage{spverbatim}
\usepackage{graphicx}
\usepackage{subfig}
\usepackage{subcaption}
\usepackage[export]{adjustbox}
\usepackage{wrapfig}
\usepackage[font=small,labelfont=bf]{caption}
\usepackage[toc,page]{appendix}
\usepackage{float}
\usepackage{amssymb}
\usepackage{amsthm}
\usepackage{chngcntr}
\usepackage[normalem]{ulem} 
\newcommand{\tx}{\text}
\usepackage[colorlinks=true, allcolors=blue]{hyperref}
\usepackage{xcolor}
\usepackage{datetime}
\usepackage[normalem]{ulem}

\counterwithout{equation}{section}

\usepackage{authblk}

\begin{document}

\title{ Microcavity-Enhanced Exciton Dynamics in Light-Harvesting Complexes: Insights from Redfield Theory}

\author[1]{Ilmari Rosenkampff\thanks{Current address: Department of Inorganic Spectroscopy, Max Planck Institute for Chemical Energy Conversion, 45470 Mülheim an der Ruhr, Germany}}
\author[1]{T\~onu Pullerits\thanks{Corresponding author: tonu.pullerits@chemphys.lu.se}}

\affil[1]{Chemical Physics and NanoLund, Lund University, Box 124, 22100 Lund, Sweden}

\maketitle  

\section*{Abstract}

We investigated the exciton transfer dynamics in photosynthetic light-harvesting complex 2 (LH2) coupled to an optical microcavity. Using computational simulations based on Redfield theory, we analyzed how microcavity coupling influences energy relaxation and transfer within and between LH2 aggregates. Our results show that the exciton transfer rate between B850 rings follows a square dependence on the light-matter coupling strength, in agreement with Fermi’s golden rule. Interestingly, the energy transfer rate remains almost independent of the number of LH2 complexes. This behavior is explained by the molecular components of the polaritonic wavefunction overlaps. These findings highlight the crucial role of cavity-induced polaritonic states in mediating energy transport and provide a theoretical framework for optimizing microcavity environments to enhance exciton mobility in light-harvesting systems and related photonic applications.

Topics

\textbf{Light-harvesting complexes, Excitons, Open quantum systems, Molecular polaritons, Optical microcavities, Quantum dynamics}

\section*{I. INTRODUCTION}

Experimental studies on strong coupling between photosynthetic structures and microcavities have been conducted on isolated light-harvesting complexes  \cite{natcomm}, organelles \cite{chlorosome}, and even on whole bacteria \cite{bacteriacoupling}. The light-harvesting complex of purple bacteria, LH2, provides a good model platform for polaritonic research not only due to its well-known structure \cite{crystalstructure,crystalanother} and spectral and  energy transfer properties \cite{tdependentlh2,reconcilinglh2dynamics, grondelleET,beforeforster}, but also because of the large transition dipole moments that promote strong coupling between the LH2 excitons and cavity modes \cite{reconcilinglh2dynamics,newdirection}. Coupling photosynthetic structures to cavities offers opportunities to alter photosynthetic energy transfer
pathways non-invasively \cite{natcomm,cavitymodified,chlorosome,bacteriacoupling}.  Furthermore, enhancing excitonic energy transfer with strong light-matter coupling may impact the improvement of organic solar cells or organic photovoltaic catalysts, whose power conversion efficiency is often limited by short exciton diffusion lengths \cite{organiccell,organiccatalyst,nonfullerenediffusion}.

When an optical cavity is close to resonant with electronic excitation of material, the electromagnetic mode of the cavity hybridizes with the material excitations, leading to Rabi splitting of the energy levels into upper and lower polaritons. Polaritons are linear combinations of electronic and photonic excitations and involve a high degree of excitonic delocalization within a molecular population. The condition of strong coupling between the cavity and the excitons requires that the excitons and the cavity mode must exchange energy faster than their decay processes. In addition to large magnitude transition dipole moments, the Rabi splitting is enhanced by a high concentration of coupled molecules within the cavity mode volume. \cite{newdirection,hybridstates}

LH2 contains bacteriochlorophyll $a$ (Bchl) and carotenoid molecules as chromophores. The Bchls in LH2 of the purple bacterium \textit{Rhodopseudomonas acidophila} form two ring structures, B800 and B850, which comprise 9 and 18 Bchls, respectively. The Bchls in the B800 ring are well separated, whereas the short spacing of the Bchls in the B850 ring induces significant excitonic coupling between the chlorophylls. Excited B850 chlorophylls form molecular, or Frenkel excitons, in which the excitation is delocalized among several Bchls and the corresponding absorption band is shifted to lower energies \cite{tonuexciton,grondelle}.  Consequently, the B850 ring and the largely non-interacting B800 ring chlorophylls absorb light at 850 nm and at 800 nm, respectively.  \cite{LH2chapter, purplebacteria}

Theoretical and experimental studies on molecular aggregates coupled to optical cavities have been performed to understand the role of the optical cavity in excitation energy transfer and conductivity, which are known to be enhanced with strong cavity-exciton coupling \cite{polaritonmediated,cavitymodified,nonradiativetransfer,extraordinarycond,cavityenhancedtransport}. Hybrid polaritonic states that are largely delocalized across the molecular population play a crucial role for the enhanced rate \cite{polaritonmediated,cavitymodified,nonradiativetransfer}. However, only a few of the states of a cavity-aggregate system are polaritonic, as the rest remain largely uncoupled, i.e. excitonic \cite{natcomm,entropy,cavitymodified}. Previous experimental findings demonstrate the importance of the dark states also during the decay of a population of LH2s strongly coupled to an optical cavity: the energy transfer is ultimately directed towards the dark states, and consequently, the lifetime of the whole system is dictated mainly by the decay of the dark states   \cite{natcomm}. Increased exciton-exciton annihilation rate in the cavity LH2 sample due to the cavity-mediated excitation transfer has been also reported. Interestingly, the effect was observed even without strong coupling regime \cite{fanwurecent}.
 
Here, we investigated cavity-mediated energy transfer between LH2s using an open quantum system approach with Redfield theory. We excited a B800 Bchl, observed its relaxation to a lower-lying B850 exciton states within the same LH2, and then calculated the energy transfer rate from the B850 exciton to other LH2s that are coupled only via a cavity mode. We found that the energy transfer rate between LH2s increases linearly with the square of the magnitude of the cavity-exciton coupling, obeying Fermi's golden rule. Surprisingly, the rate did not increase with the number of LH2s, which we ascribed to the scaling of the wavefunction overlap between two B850s with $1/N_\text{LH2}$. With cavity-exciton coupling of 0.2 cm$^{-1}$, 75\% of the B850 to B850 energy transfer was mediated through polaritons, and a smaller fraction of 25\% accounted for direct B850 to B850 transfer. At couplings above $\sim$0.6 cm$^{-1}$, direct B850 to B850 energy transfer was favored.

\section*{II. THEORY}

\subsection*{A. The Hamiltonian}

Assuming a single cavity mode of a Fabry-P\'erot cavity coupled with molecular excitons, the polariton Hamiltonian \cite{jaynescummings,taviscummings} can be written in the form 

\begin{equation} \label{hamiltonian}
\begin{split}
        \tx{H}= \sum_{i} (E_{i} + \Delta_{i})b_i^\ddag b_i +  \sum_{i \neq j}V_{ij} (b_i^\ddag b_j +b_i b_j^\ddag)  )\\ + \omega_C a^\ddag a +   \sum_i  g_i (b_i a^\ddag + b_i^\ddag a), 
\end{split}
\end{equation}
where $\omega_C$ is the energy of the cavity mode with the photon annihilation and creation operators $a$ and $a^\ddag$, respectively. The corresponding operators for the molecular excitations are $b^\ddag$ and $b$ \cite{newdirection,nonhermitian}. The Bchl excitation energies are \textit{E\textsubscript{i }}, and $\Delta$\textit{\textsubscript{i }}represent random shifts of the transition energies due to the fluctuations in the protein environment, also called inhomogeneous broadening. \textit{V\textsubscript{ij }}are the dipole-dipole couplings between Bchl \textit{i }and \textit{j }transition dipole moments   \cite{cavitymodified,tonupaper,reconcilinglh2dynamics}. The static energy disorders $\Delta$\textit{\textsubscript{i }}were assigned pseudo-randomly, ensuring that within a given LH2 complex, the disorder values remained consistent across different simulations, while varying between different LH2 complexes. These values were drawn from a Gaussian distribution with a full width at half maximum (FWHM) of 93 cm\textsuperscript{-1 }for the B800 chlorophyll and 160 cm\textsuperscript{-1 }for the B850 chlorophylls, following \cite{scholes}.

The mode of the Fabry-P\'erot cavity interacts with the molecular excitations via the coupling elements
\begin{equation}\label{couplingeqn}
    g_i  = - g_0 \boldsymbol{\mu}_i \cdot \textbf{u},
\end{equation}
where \textit{\textbf{µ}\textsubscript{i }} is the transition dipole moment of the Bchl \textit{i}, and \textbf{u }is the unit vector in the direction of the field polarization \cite{nonhermitian,newdirection}. With spin coated LH2 films, the LH2s have a tendency to orient so that the two rings of Bchl molecules are roughly parallel with the cavity plane  \cite{spincoated}. In this way the transition dipole moments of the Bchls are all in the plain of the cavity. We considered only the cavity field x component and took the light mater interactions of each Bchl molecule to be proportional of their transition dipole moment x component. 

Since we targeted dynamics between the complexes, we concentrated on B850 Bchls. One B800 Bchl was included in one of the LH2s to provide a starting point for simulations as a B800 molecular excitation. 18 B850 Bhls of all LH2s were included in the Hamiltonian. Couplings of Bchls belonging to different LH2s were set to zero. The polaritonic Hamiltonian was diagonalized, giving eigenstates of the form

\begin{equation}\label{polariton}
        \ket{J}= c_J(0) \ket{1_c,G} + \sum_{i} c_J(i) \ket{0_c, i},
\end{equation}
where $\ket{0_c,i}$ are the molecular site excitations with unoccupied cavity mode, and $\ket{1_c,G}$ is the photon-occupied cavity mode with the LH2s in the ground state. The cavity energy $\omega_C$ was set to 11765 cm$^{-1}$ (850 nm). Photonic and B850 Hopfield coefficients of states were defined as

\begin{align}
    \eta_{photonic}(J) = |c_J(0)|^2, \\
    \eta_\text{B850}(J)= \sum_{i\in \mathrm{B}850} |c_J(i)|^2,
\end{align}
where $c_J(0)$ is the probability amplitude of the cavity mode \cite{purplebacteria}, and the sum over $i$ includes amplitudes of B850 Bchls belonging to the B850, or groups of B850s, in question. We defined polaritons as states with a photonic Hopfield coefficient greater than 0.1 and B850 states as states with B850 Hopfield coefficient greater than 0.9. Simulations without the cavity were carried out using only the molecular part, or the first two sums of the Hamiltonian in Equation \ref{hamiltonian}, yielding excitonic eigenstates
\begin{equation}
    \ket{\alpha} = \sum_i c_\alpha(i) \ket{i},
\end{equation}
which are superpositions of the excited molecular sites $\ket{i}$ \cite{classicalexcitonpaper,tonupaper}.

\subsection*{B. Relaxation model}

Polariton relaxation was simulated with Redfield theory in which the relaxation is mediated by coupling of the cavity coupled states to the vibrational modes of the protein scaffold. The approach is outlined in  \cite{tonupaper,classicalfluorescencedynamics} but rephrased here so that the coupling to the cavity is included as in   \cite{photoluminescence}. The equations of motion for the population of the state $J$ can be written as
\begin{equation} \label{master}
     \dot{p}_J = \sum_{J'} (W_{J'J} p_{J'} - W_{J J'} p_J ),
\end{equation}
where $W_{J'J}$ is the transfer rate from the state $\ket{J'}$ to $\ket{J}$. We took \textit{W\textsubscript{JJ'}} = 0 for \textit{J }= \textit{J}' since we targeted the dynamics which is much faster than the excited state lifetime. The finite excited state lifetime was only included in simulations of a plain LH2 (c.f. Equation \ref{ratematrix}, and Figure \ref{LH2internalrelaxation}).

Inter-state energy transfer occurs via the coupling between the molecular states and nuclear motions (diagonal coupling). The transfer rates can be written as
\begin{equation} \label{wrates}
     W_{J J'} = \begin{cases}
     \frac{2\pi}{\hbar}(1+ n(-E_{J J'},T))D(-E_{J J'}) \sum_i |c_{J}(i)|^2 |c_{J'}(i)|^2 \qquad E_{J J'}<0, \\
     \frac{2\pi}{\hbar} n(E_{J J'},T))D(E_{J J'}) \sum_i |c_{J}(i)|^2 |c_{J'}(i)|^2 \qquad E_{J J'}>0,
     \end{cases}
 \end{equation}
where $n(E_{JJ'},T)$ is the Bose-Einstein distribution evaluated with the energy difference between the states J and J', $E_{J J'}= E_J' - E_{J}$, and the sum is carried over all Bchl sites $i$.  The first equation describes the downwards rate consisting of both stimulated and spontaneous components while the second equation is for the uphill rate that can only occur via a stimulated process where vibrations give energy to the electronic system.

\textit{D}(\textit{E\textsubscript{JJ}}') is the spectral density of the electron-phonon coupling,
\begin{equation} \label{amplitudej}
    D(E_{J J'}) = j_0 d(E_{J J'}),
\end{equation}
where the amplitude \textit{j}\textsubscript{0 }depends on the coupling strength between the system and the bath. The Huang-Rhys factor weighted density of phonon states multiplied by a square of energy \textbf{was} taken to be \begin{equation}
     d(E) = \frac{E^2}{E_0^3} \exp(-\frac{E}{E_0}),
 \end{equation}
where the parameter \textit{E}\textsubscript{0} was set to 100 cm\textsuperscript{-1}, adjusting the maximum of \textit{D}(\textit{E}) to the thermal energy of about 200 cm\textsuperscript{-1}. For obtaining the spectral density from various optical experiments and calculations, we refer the interested readers to the following literature \cite{pulleritsspectraldensity,pulleritsdensity2,environmenttheory}. Importantly, the energy relaxation rate is a function of the overlap of the molecular wavefunctions between the states represented via the sums over the squared wavefunction amplitudes, henceforth referred to as overlap sums. 

The system of equations of motion given by Eq. \ref{master} can be cast in a matrix form,
\begin{equation}
    \dot{\textbf{p}}= \text{R} \textbf{p},
\end{equation}
where the elements of  $\tx{R}$ are
\begin{align}
\label{ratematrix}
    \tx{R}_{\tx{JJ}'} = - \sum_{\tx{J}''}W_{\tx{JJ}''}- 1 /\tau_\tx{exc}, \quad  \tx{J} = \tx{J'},\\
    \tx{R}_{\tx{JJ}'} = W_{\tx{J'J}}, \quad  \tx{J} \neq \tx{J'}.
\end{align}
The solution is then given as 
\begin{equation}
    \textbf{p}(t)= \sum_i h_i e^{\lambda_i t} \textbf{m}_i,
\end{equation}
where $\textbf{u}_i$ and $\lambda_i$ are the eigenvectors and corresponding eigenvalues of the matrix $\tx{R}$. The coefficients  $c_i$ were solved via
\begin{equation}
    \textbf{h}= \text{M}^{-1} \textbf{p}(0),
\end{equation}
where the matrix M is such that M$_{ij}=[\textbf{m}_i]_j$, and the vector $\textbf{h}$ contains the coefficients $h_i$. For solving the equations of motion, the initially populated state was chosen to be an excitonic state localized at the B800 Bchl with an energy of approximately 800 nm, with probability amplitude squared of the B800 Bchl of at least 0.5. Excited-state lifetime $\tau_\text{exc}$ of 200 ps was used, corresponding roughly to the cavity dark state lifetime of the LH2 from the purple bacterium \textit{Rhodoblastus acidophilus} \cite{natcomm}. 

\section*{III. RESULTS AND DISCUSSION}

\subsection*{A. Exciton relaxation in LH2}
\label{excitonrelaxation}

\begin{figure}[H]
    \centering
    \includegraphics[width=\textwidth]{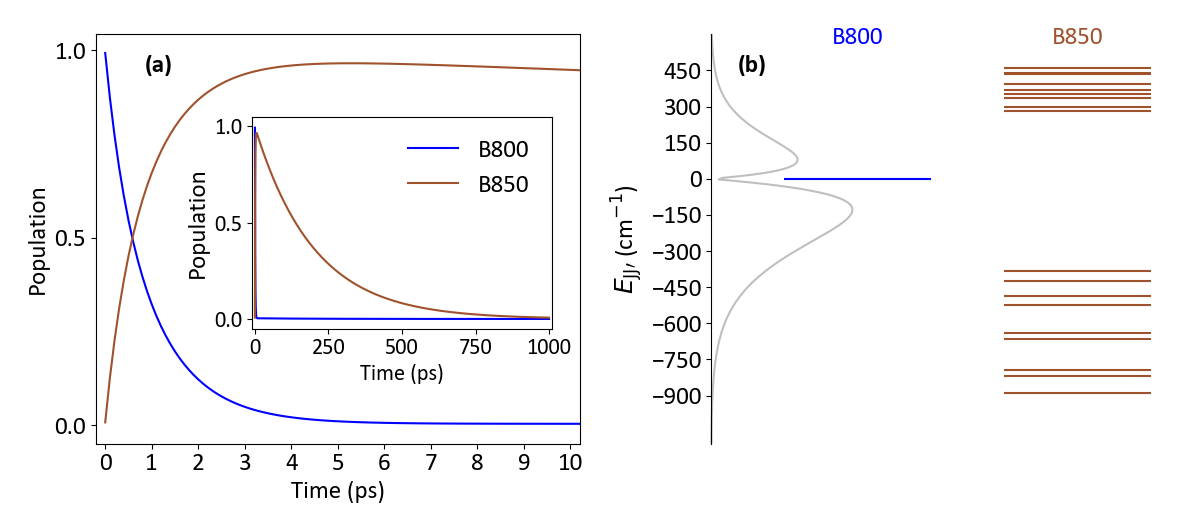}
    \caption{\textbf{(a) }Population decay of the initially excited B800 (blue) and the corresponding excitation population rise of the B850. The inset shows long timescale dynamics. (\textbf{b}) The energy level structure of LH2 showing a single B800 level as blue (we only \textbf{included} one B800 Bchl) and 18 B850 exciton levels, beige. On the left, the grey plot is the temperature weighted spectral density function used in Equation \ref{wrates} where both the upwards and downwards rate contributions are included.  }
    \label{LH2internalrelaxation}
\end{figure}

Exciton relaxation was first simulated in bare LH2 as in \cite{tonupaper}. The B800- and B850-exciton populations were defined as
\begin{equation} \label{b800pop}
    p_\mathrm{B800}(t)=\sum_{\alpha, i\in \mathrm{B}800} |c_\alpha(i)|^2 p_\alpha(t) 
\end{equation}
\begin{equation} \label{b850pop}
    p_\mathrm{B850}(t)=\sum_{\alpha, i\in \mathrm{B}850} |c_\alpha(i)|^2 p_\alpha(t),
\end{equation}
where the sum over $\alpha$ accounts for all excitonic states of the LH2, and the sum over \textit{i }includes either the amplitude of the B800 Bchl or the amplitudes of the 18 B850 Bchls. The B800$\rightarrow$B850 energy transfer time is 0.9 ps in \textit{Rhodoblastus acidophilus} \cite{0.7pstransfer}. The electron-phonon coupling strength parameter \textit{j}\textsubscript{0 }was adjusted so that the calculated B800$\rightarrow$B850 transfer time matched $\sim$0.9 ps, see Figure \ref{LH2internalrelaxation}. For the following cavity calculations, the same LH2 parameters were used.

\subsection*{B. Coupling to the cavity}

\begin{figure}[H]
    \centering
    \includegraphics[width=\textwidth]{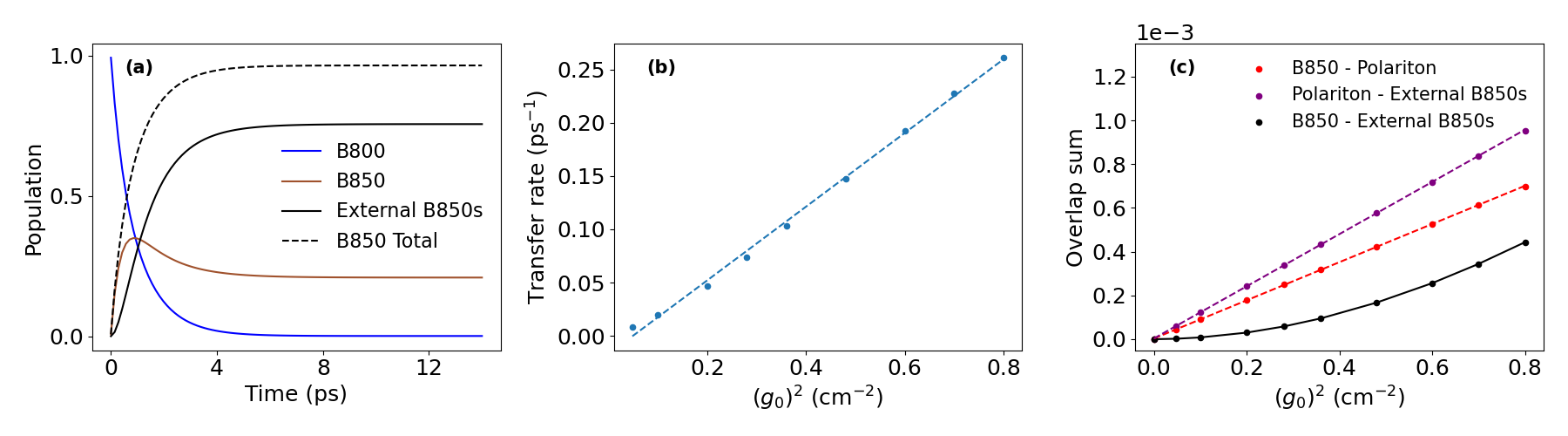}
    \caption{\textbf{(a)} Relaxation simulation with a single LH2 containing an initially excited B800 and the B850 ring, and three external B850 rings each with 18 Bchls;  \textbf{(b)} relaxation rate from the initially excited LH2 to the external B850s as a function of square of the coupling; \textbf{(c)}: overlap sums between the initially excited B850 and the polariton (red), the polariton and the external B850s (purple), and the initially excited B850 and the external B850s (black) as a function of the square of the coupling. Coupling constant $g_0$ in the simulation in \textbf{(a)} is set to 2 cm$^{-1}$. Transfer rate for each coupling in \textbf{(b)} was obtained by fitting the initial rise of the population of the external B850s with a single exponential. The overlap sums in \textbf{(c)} were calculated as described in the main text, and the populations in \textbf{(a)} similarly as in Equations \ref{b800pop}-\ref{b850pop}. Dashed lines are linear fits.}
    \label{cavitymediated}
\end{figure}

Cavity-mediated energy transfer was simulated in a system of one initially excited LH2 and three additional external B850 rings (Figure \ref{cavitymediated}). The cavity-exciton coupling parameter $g_0$ was varied, and the simulated external B850 occupancy was fitted with an exponential with the method of least squares. The energy transfer rate from the B850 belonging to the initially excited LH2 to the external B850s was plotted against the square of the coupling strength. In lower limits of the coupling, B800 to B850 transfer occurs before the transfer to the rest of the B850s (Figure \ref{cavitymediated}\textbf{a}). Increasing the coupling strength to the cavity in this regime increases the rate of energy transfer to the external B850 rings linearly as a function of the square of the cavity-exciton coupling, in accordance with the usual Fermi’s Golden rule (Figure \ref{cavitymediated}\textbf{b}) even though the coupling is not direct but involves the cavity as mediator. Still, the coupling mediated by the cavity leads to the inter-complex energy transfer similarly to the  excitation transfer between directly coupled systems outside the cavity.

The energy transfer rate from B800 to the total B850 manifold remains unaffected by the cavity in this range of the couplings (Figure \ref{cavitymediated}\textbf{a}), while the cavity is redistributing excitations between the B850s. The energy transfer from B850 to external B850s can take two possible routes: either directly from the initially excited B850 to the external B850 rings, or indirectly from the initially excited B850 to the polariton, and from the polariton to the external B850s. The two lowest energy B850 states of the initially excited LH2 comprise 50\% of the total population of all the B850 states (Figure \ref{b850statepopulations}). Furthermore, transfer time from the B850 to external B850 is about 4 ps at ($g_0$)$^2$ = 0.8 cm$^{-2}$ $-$ clearly slower than the subpicosecond transfer from the high energy B800 state to the B850 states (Figure \ref{cavitymediated}). Thereby, the energy transfer occurs mainly through the two lowest B850 states.

To investigate the different energy transfer pathways, overlap sums between the relevant groups of states were evaluated. The overlap sum between B850 and the polariton was calculated as the sum of two overlap sums between the two low energy B850 states and the single polariton state (Figure \ref{cavitymediated}\textbf{c}). The overlap sum between B850 and the external B850s was calculated similarly, but by only considering the external B850 states with energies lower than the highest of the two above mentioned initial B850 states (11719 cm$^{-1}$) (Figure \ref{energiesvsg0}). For calculating the overlap sums, external and initially excited B850 states were defined as states with the corresponding (external or initially excited) Hopfield coefficient greater than 0.9. 

 The calculated overlap sums favor energy transfer via the polariton: not only are the overlap sums larger between the polariton and the B850 states than between the B850 and the external B850 states, but also the polaritonic sums scale linearly with the square of the coupling, matching a similar trend of the transfer rates (Figure \ref{cavitymediated}\textbf{c}). Interestingly, linear scaling is also observed in the B850 Hopfield coefficient of the polariton (Figure \ref{cavityvsb850externalcharacter}\textbf{c}). The overlap between the B850 and external B850 wavefunctions follows a higher-than-quadratic dependence. This is expected because there is no direct interaction between the LH2s; coupling is mediated by the cavity. In perturbation theory, this implies second-order matrix elements of the form $|\braket{f|\hat{g}|c} \braket{c|\hat{g}|i}|^2$, where $\hat{g}$ is the interaction operator between the molecular state $i$ or $f$ and the cavity mode $c$.

At higher couplings, even the overlap sum between the B850 and the external B850s starts increasing with ($g_0$)$^2$ (Figure \ref{cavityvsb850externalcharacter}\textbf{a}). This matches the behavior of the external B850 Hopfield coefficient of initially excited B850 states (Figure \ref{cavityvsb850externalcharacter}\textbf{b}). The photonic Hopfield coefficient of the initially excited B850 states also scales linearly with ($g_0$)$^2$ (Figure \ref{cavityvsb850externalcharacter}\textbf{b}). Thus, increasing the light-mater interaction leads to mixing of the cavity states into the largely excitonic B850 states, until the Fermi's golden rule with respect to direct B850 to B850 energy transfer rate is followed. At couplings greater than ($g_0$)$^2$ $\approx$ 0.8 cm$^{-2}$, direct B850 to B850 transfer dominates over the polariton pathway because of the noticeably larger B850-B850 overlap compared to the B850-polariton overlap (Figure \ref{cavityvsb850externalcharacter}\textbf{a}).

\begin{figure}[H]
    \centering
    \includegraphics[width=\textwidth]{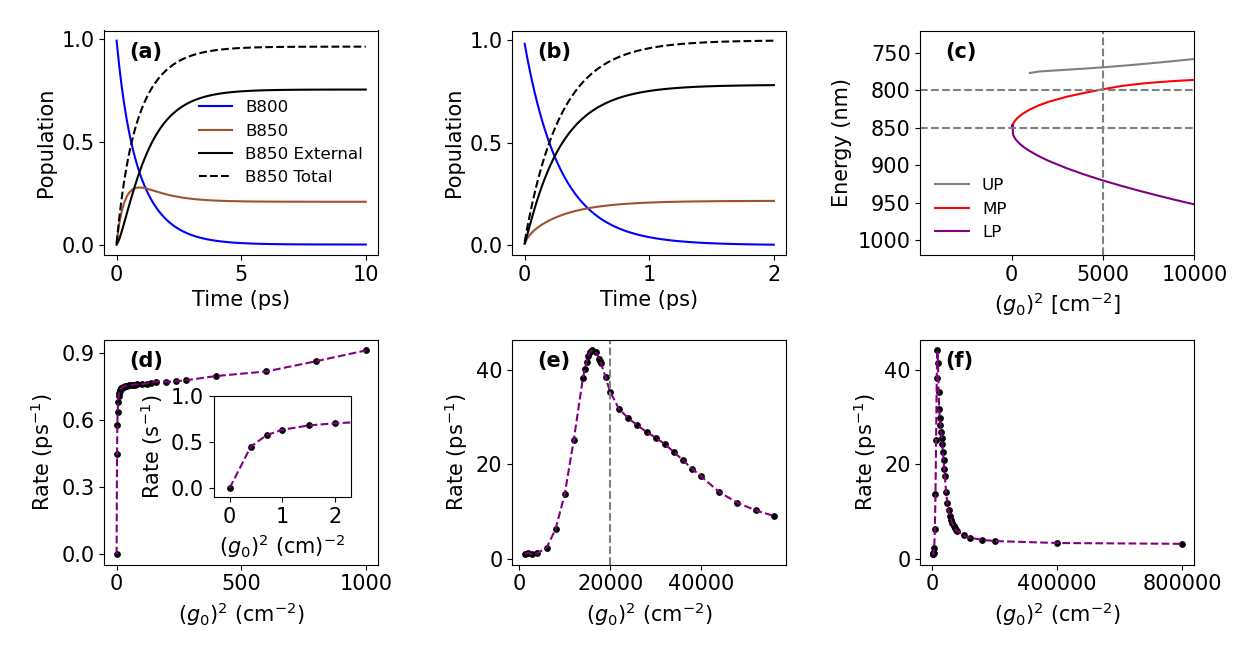}
    \caption{Relaxation simulations using a single LH2 and three extra B850 rings (B850 External) with ($g_0$)$^2$ = 80 cm$^{-2}$  (\textbf{a}) and ($g_0$)$^2$ = 400000 cm$^{-2}$ (\textbf{b}); positions of LP, MP, and UP maxima as a function of coupling (\textbf{c}); and B850$\rightarrow$B850 transfer rates as a function of coupling (\textbf{d}-\textbf{f}). The B800 level becomes resonant with the MP at 20000 cm$^{-2}$  (vertical dashed grey line). Transfer rate for each coupling was obtained by fitting the initial rise of the external B850 population plot with a single exponential. The maximum in (\textbf{e}) is at 16000 cm$^{-2}$. The energy of the cavity was set to 850 nm.
    }
    \label{fermiplots}
\end{figure}

When coupling is increased even further, the relaxation rate increase suddenly slows down when the transfer time approaches the B800 to B850 relaxation time 0.9 ps (Figure \ref{fermiplots}\textbf{a} and \textbf{d}). In such conditions, the cavity-mediated energy transfer to other B850s is limited by the transfer from B800. If coupling is increased even further, Rabi splitting of the polaritons increases, speeding up the transfer from the initially excited B800  as the middle polariton branch (MP) gradually becomes resonant with the B800 level (Figure \ref{fermiplots}\textbf{e}). Under these conditions, energy can transfer directly from the MP polaritons to external B850 rings. The largely delocalized nature of the polaritons in and of itself can directly pass the excitation to other molecular levels.

Careful analysis of the energies of the states involved in the formation of the MP branch is needed for understanding the transfer rate maximum at 16000 cm$^{-2}$ and the shoulder at 32000 cm-2. At ($g_0$)$^2$ = 16000 cm$^{-2}$, the B800 state has become increasingly polaritonic and is separated from the MP polariton by about 110 cm$^{-1}$ (Figure \ref{figs4}). Such an energy separation roughly matches the distance from the center of the spectral density function to the lower energy side maximum  (Figure \ref{LH2internalrelaxation}\textbf{b}). Thus, the likely energy transfer pathway is from the formal B800 state to the formal MP polariton and further to the lower external B850 states, which are well within reach of the lower energy tail of the spectral density (Figure \ref{LH2internalrelaxation}\textbf{b}).  At ($g_0$)$^2$ = 32000 cm$^{-2}$ in turn, the MP polariton, which is now 140 cm$^{-1}$ above the B800 state, is only about 70 cm$^{-1}$ away from the higher energy external B850 states. Thus, at this coupling, energy transfer from the polaritonic B800 to the higher B850 external states contributes to the total transfer rate.  The rate gradually decreases after about ($g_0$)$^2$ = 40000 cm$^{-2}$ as the B800 state becomes more and more molecular when the MP branch passes to even higher energies with an even greater Rabi splitting, decreasing the level of mixing between B800 and the polariton (Figure \ref{fermiplots}\textbf{e}). A limiting transfer rate of about 3 ps$^{-1}$ is reached at strongest couplings (Figure \ref{fermiplots}\textbf{f}). As a comment, we point out that such very strong couplings are far beyond what is experimentally achievable today. At the same time quite large Rabi splitting can be achieved since the splitting is proportional to the square root of the number of molecules in the cavity mode volume. This number can be very large.

Cavity-mediated coupling has been demonstrated to lead to energy transfer between spatially well-separated systems \cite{energytransferSpatiallySeparated,directevidence}. When an excited B850 transfers energy to another excited B850, a doubly excited state is formed, followed by a fast internal conversion back to a singly excited B850 state \cite{excitonannihilation}. As a result, one excitation disappears, leading to additional fast component to the excitation decay. The process is called exciton-exciton annihilation and requires high light intensities, where multiple LH2s are excited concomitantly \cite{fanwurecent}. Since the process depends on energy transfer, excitation intensity-dependent studies can be used to analyze the energy transfer dynamics and system connectivity \cite{invitrotonu}. The simulations presented herein are applicable to any type of energy transfer. It should be noted that experimentally observed exciton-exciton annihilation rates of LH2 cavity systems are at least order of magnitude slower than B800 to B850 relaxation time \cite{fanwurecent}.

\subsection*{C. Number of coupled LH2 complexes}

\begin{figure}[H]
    \centering
    \includegraphics[width=\textwidth]{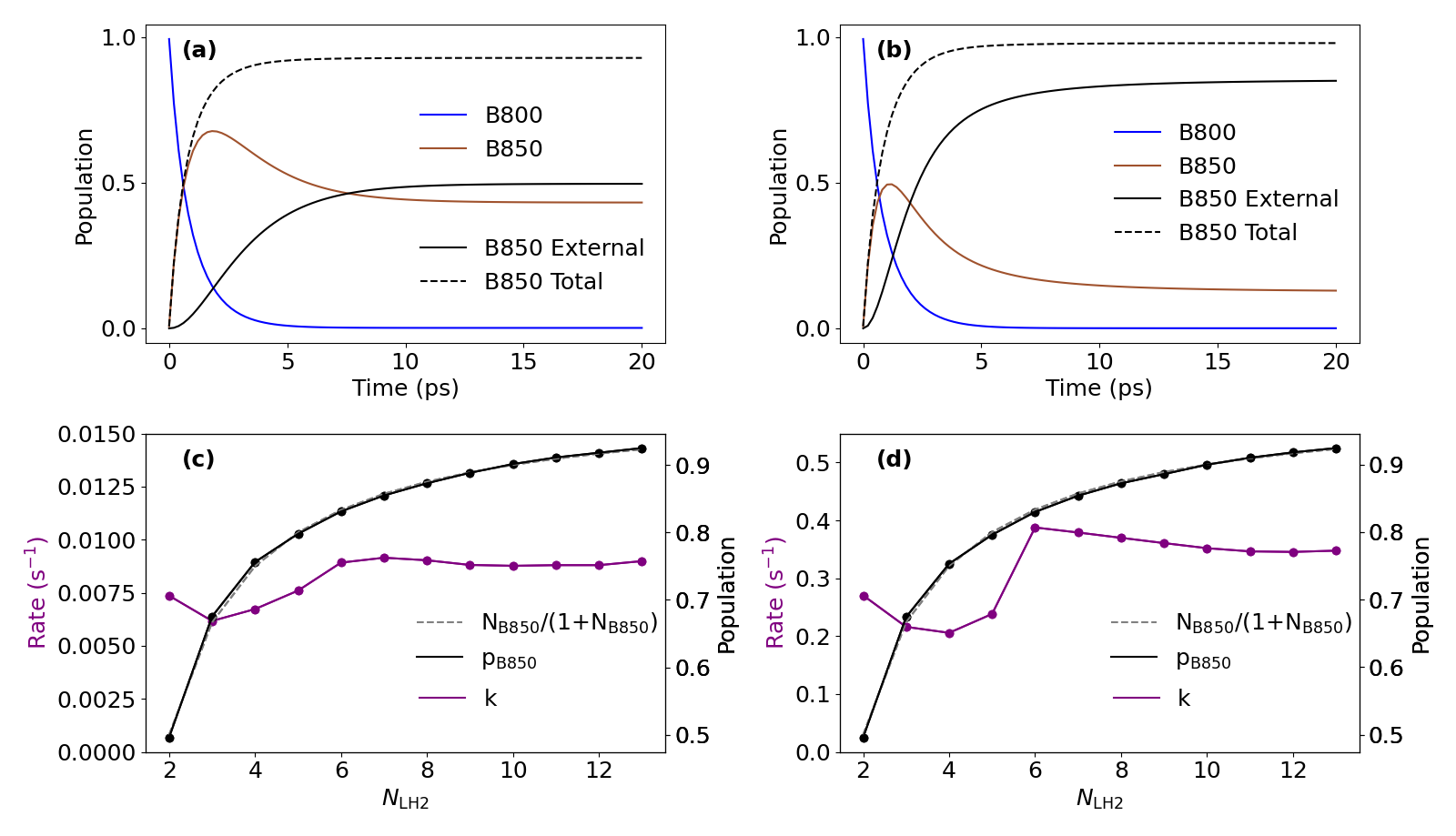}
    \caption{ Relaxation simulations with two (\textbf{a}) and seven (\textbf{b}) LH2s, i.e. with one and six external B850s, respectively, using $g_0$ = 0.8 cm$^{-1}$ in both simulations; and B850$\rightarrow$B850 transfer rate $k$ as a function of the number of LH2s with $g_0 = 0.2$ cm$^{-1}$ (\textbf{c}) and $g_0 =$ 0.8 cm$^{-1}$ (\textbf{d}), showing also the external B850 population (p$_{850}$) in steady state.  Transfer rate $k$ was evaluated by fitting the initial rise of the external B850 population with a single exponential. $N_\text{B850}$ is the number of external B850 rings.
    }
    \label{nb850}
\end{figure}

Relaxation simulations were carried out by varying the number of external B850s, as shown in Figure \ref{nb850}.We carried out the calculations with two different couplings of 0.2 and 0.8 cm$^{-1}$, leading to the energy transfer times of about 100 ps and 3 ps, respectively. For comparison, the cavity-mediated energy transfer times that can be deducted from annihilation experiments are a few tens of picoseconds \cite{fanwurecent}. Clearly, the calculated energy transfer rate does not significantly depend on the number of LH2s, which is at first glance surprising since the cavity connects all LH2s and the larger number of accepting systems should mean higher rate.

The population of the external B850 manifold always increases with the number of external B850 rings, as energy is distributed across a larger number of complexes. Specifically, the external B850 population closely follows the expected $N_\text{B850}/(1 + N_\text{B850})$ in a steady state, where $N_\text{B850}$ is the number of external rings, implying that the few polaritonic states do not contribute significantly to the total occupation.

\begin{figure}[H]
    \centering
    \includegraphics[width=\linewidth]{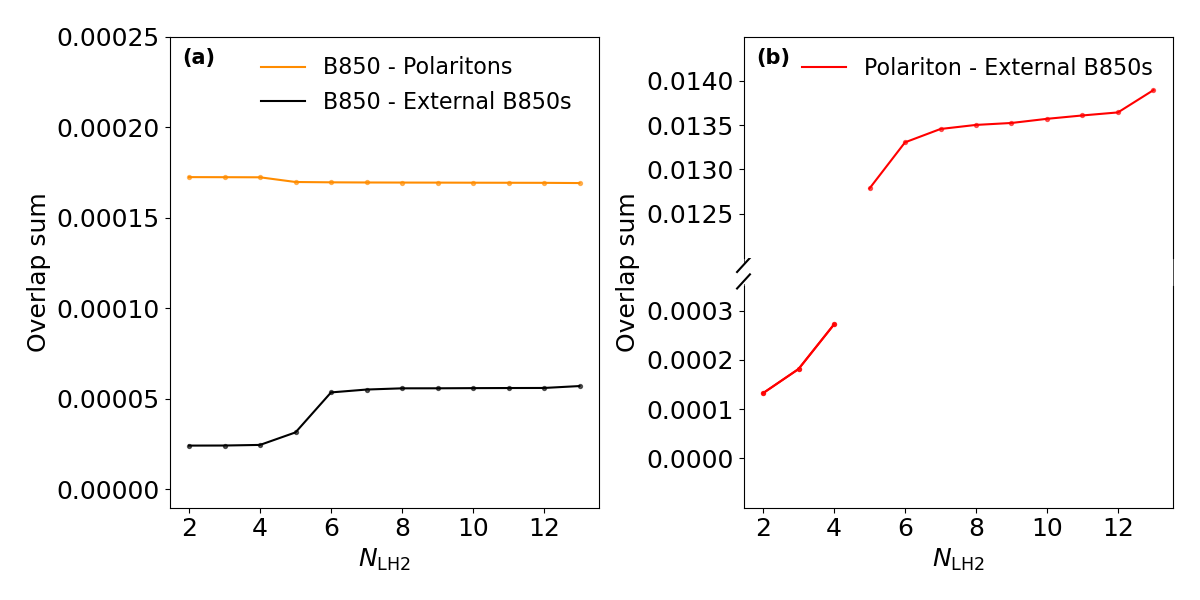}
    \caption{ \textbf{(a)} Wavefunction overlap between the initially excited B850 and polaritons (orange), and between the initially excited B850 and the external B850s (black); \textbf{(b)} wavefunction overlap between the polaritons and the external B850s. An external B850 state close to resonance with the polariton is present with $N_\text{LH2}$ $>$ 4.  The overlaps were calculated as described in the main text with $g_0$=0.2 cm$^{-1}$.}
    \label{mainoverlaps}
\end{figure}

To understand the very week dependence of the energy transfer on  $N_\text{LH2}$, the overlap sums were calculated between the initially excited B850, external B850s, and polaritons similarly as before, but by including each of the initially excited B850 and external B850 states into the calculation and using $g_0$ = 0.2 cm$^{-1}$ (Figure \ref{mainoverlaps}). Despite increasing number of external B850 states, significant wavefunction overlap between the polaritons and the initially excited B850 states favors energy transfer via the polariton to the external B850 states. Overlap between the polaritons and the external B850 states increases drastically with the number of LH2s, with a sharp rise when going from four to five LH2s (Figure \ref{mainoverlaps}\textbf{b}). This sharp increase is due to the appearance of an external B850 state, which is nearly resonant with the polariton (Figure \ref{energiesvsnlh2}). 

An important observation about the mixing of dark states is that, unlike the mixing between polaritons and dark states, the mixing among dark states does not increase drastically even when new dark states close to resonance with the existing ones are introduced into the system. However, the jump in the rate when going from 5 to 6 LH2s is due to an external B850 state that appears only $\sim$0.03 cm$^{-1}$ from the initially excited B850 state (Figure \ref{energiesvsnlh2}). In real experiments, millions of molecules can be involved  \cite{millionofmolecules}, and appearance of such near resonances are quite likely to happen. On the other hand, including realistic linewidths to the state energies would smoothen such spurious behavior \cite{plexcitonpaper}. Using a higher coupling of 0.8 cm$^{-1}$ makes the jump in the rate from 5 to 6 LH2s more significant (Figure \ref{nb850}\textbf{d}), indicating an increased dark state mixing with $g_0$. Mixing between dark states farther apart from each other, however, does not appear to increase with stronger coupling to the cavity.

Analytical solution of Equation \ref{master} with the initial condition $p_\text{B850}(0) = 1$ provides an energy transfer rate from the initially excited B850 as 
\begin{equation}
    k=k_\text{ext}N_\text{LH2} + k_\text{c},
\end{equation}
where \textit{k}\textsubscript{ext }is the transfer rate constant between two B850s, and \textit{k}\textsubscript{c }is the B850→polariton transfer rate (see Appendix \textbf{A }for details). In our case clearly \textit{k}\textsubscript{c} is the rate-limiting step. According to Figure \ref{mainoverlaps}\textbf{a}, the overlap between the initial B850 and the polariton is slightly decreasing. This qualitatively agrees with the prediction of the exciton to polariton transfer rate behavior \cite{photoluminescence,jaggregatedynamics}. If the number of complexes increases, the molecular part of the polariton state is shared by a large number of mater states, which means that the contribution (the overlap) by a certain complex (our initially excited LH2) is expected to decrease. On the other hand, the overlap between a polariton and the external B850s increases on the basis of the same logic. Indeed, the calculated overlap drastically increases with \textit{N}\textsubscript{LH2}. The two trends partially compensate, and the net result for the two-step process of transfer from a B850 to a polariton and from there to an external B850, is a weak dependence on the number of complexes.

The direct transfer overlap between the initially excited B850 and each of the external B850s does not change significantly with the number of external complexes for the parameter set that leads to realistic transfer times (Figure \ref{mainoverlaps}\textbf{a}). The B850 - Polariton overlap is about 3 times greater than the total B850 - external B850 overlap. This suggests that about 25\% of the transfer from the initially excited B850 to the external B850s occurs via the direct molecular pathway, and 75\% proceeds via the polaritons. It should be noted that the cavity energy in the simulations was higher than that of the lowest-lying excitons, which may influence the energy-transfer pathways.  As a comment, we point out that annihilation experiments with 0.5, 1.0 and 1.5 lambda cavities, where the concentration of the molecules was kept constant, have shown that cavity-mediated transfer depends much more on $g_0$ than on $N$, supporting the findings of our calculations.

\section*{IV. CONCLUSIONS}

Our simulations suggest that cavity-mediated excitation energy transfer between molecular B850 units is largely independent of their number. Polaritons act as intermediate energy acceptors, facilitating efficient excitation energy transfer from the donor molecule (B850) to the acceptors (external B850s). We found the rate of such polariton-assisted energy transfer to scale with the square of the coupling to the cavity. Under very strong coupling, direct energy transfer from the initially excited molecule (either B800 or B850) to the energy acceptors (external B850s) dominates. While our simulations are limited to a relatively small number of molecules, in real experiments, optical microcavities are known to typically enclose millions of molecules. These results could be utilized in future designs of polariton-supporting substrates.

\section*{ACKNOWLEDGMENTS}

This work was supported by Swedish Energy Agency grant 50709-1, Swedish Research Council grant 2021-05207 and Olle Engkvist foundation grant 235-0422. We also thank Dr. Fan Wu for valuable discussions.

\section*{AUTHOR DECLARATIONS}

\subsection*{Conflict of Interest}

The authors have no conflicts to disclose.

\subsection*{Author Contributions}

\textbf{Ilmari Rosenkampff}:  Formal analysis (equal); Investigation (equal); Writing – original draft (lead); Writing – review \& editing (supporting). \textbf{Tõnu Pullerits}: Conceptualization (lead); Formal analysis (equal); Investigation (equal); Supervision (lead); Writing – original draft (supporting); Writing – review \& editing (lead).

\section*{DATA AVAILABILITY}

The data that support the findings of this study are available from the corresponding author upon reasonable request.

\bibliographystyle{jcp}

\section*{Supplementary information}
\label{SI}

\newcounter{Sfig}
\renewcommand{\thefigure}{S\arabic{Sfig}} 

\newcommand{\SIFigure}[4][]{ 
    \refstepcounter{Sfig} 
    \begin{figure}[H]
        \centering
        \includegraphics[width=#1]{#2}  
        \caption{#3} 
        \label{#4}  
    \end{figure}
}


\SIFigure[\textwidth]{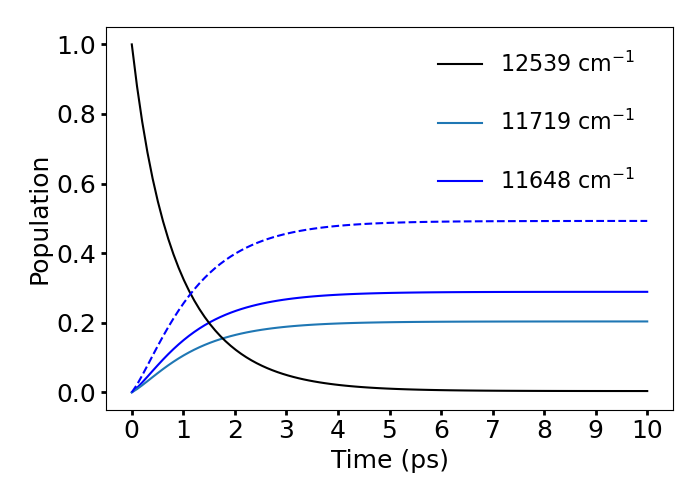}{Populations of the two lowest energy B850 states (blue and light blue curves), their sum (dashed line), and the population of the initially excited B800 state (black). The simulation includes a single LH2 with one B800 Bchl and 18 B850 Bchls.}{b850statepopulations}

\SIFigure[\textwidth]{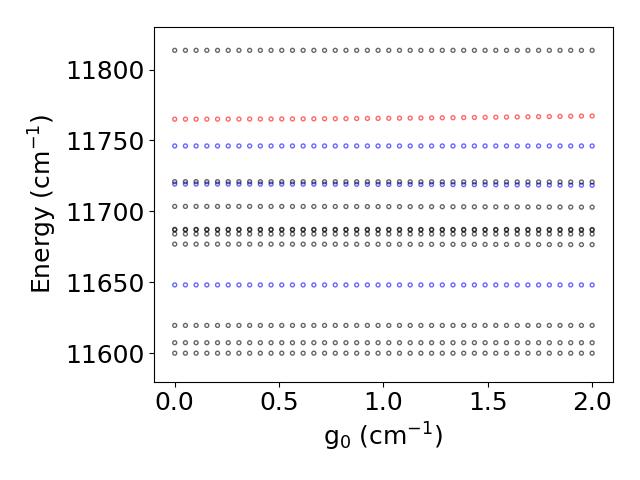}{Lowest state energies dominated either by B850 (blue), external B850 (black), or polariton (red) character as a function of coupling g$_0$ with 4 LH2s. State was defined to be a B850 state or external B850 state if the sum of the squared probability amplitudes, or the Hopfield coefficient of the B850 or external B850 Bchls was more than 0.9; state was defined to be a polariton if the squared probability amplitude of the cavity mode was higher than 0.1.}{energiesvsg0}

\SIFigure[\textwidth]{FigureS3.png}{(\textbf{a}) Overlap sums between the initially excited B850 and the polariton (red), the polariton and the external B850s (purple), and the initially excited B850 and the external B850s (black) as a function of the square of the coupling with 4 LH2s. (\textbf{b}) Photonic and external B850 Hopfield coefficient (HC) of the two lowest energy B850 states (yellow and black, respectively). (\textbf{c}) B850 Hopfield coefficient of the polariton.}{cavityvsb850externalcharacter}

\SIFigure[\textwidth]{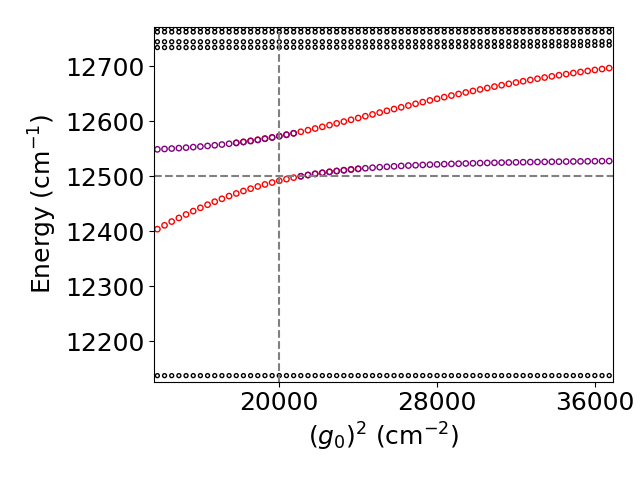}{Energies of the states involved in the MP branch as a function of coupling squared with 4 LH2s when the B800 level (12500 cm$^{-1}$) is close to resonant with the relevant polariton.  B800 states (purple) are those with probability amplitude squared of the B800 Bchl greater than 0.5; external B850 states (black) have the sum of the squared probability amplitudes of external B850 Bchls greater than 0.8; polaritons (red) have the squared probability amplitude of the cavity mode greater than 0.1. The vertical dashed line indicates the coupling at which the B800 level is resonant with the MP.}{figs4}

\SIFigure[\textwidth]{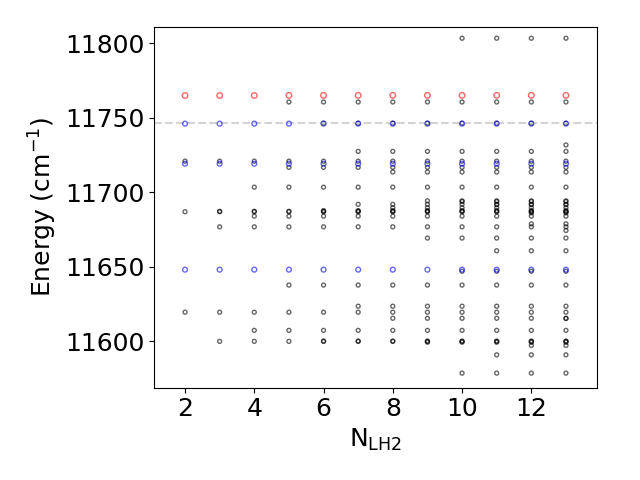}{Lowest states dominated either by B850 (blue), external B850 (black), or polariton (red) character as a function of the number of LH2s with $g_0 = 0.2$ cm$^{-1}$. State was defined to be a B850 state or an external B850 state if the sum of the squared probability amplitudes, or the Hopfield coefficient of the B850 or external B850 Bchls was more than 0.9; state was defined to be a polariton if the squared probability amplitude of the cavity mode was greater than 0.1. The horizontal dashed line indicates an external B850 state that appears at N$_\text{LH2}$=6 very close (11746.42 cm$^{-1}$) to resonance with the B850 state of a similar energy (11746.12 cm$^{-1}$).}{energiesvsnlh2}

\section*{Appendix A}
\label{appendix}

\textbf{Calculation of the energy transfer rate constant}

Let $p_\text{B850}$ be the population of the B850 ring,  $p_\text{c}$ be the population of the polaritons, and $p_\text{ext}$ be the population of a single external B850 ring. The energy transfer rate defining equation (Equation \ref{master}) takes the form
\begin{align}\label{b850transfereq}
   \frac{d}{dt} p_\text{B850} = - p_\text{B850} (k_\text{ext}N_\text{ext}   +k_\text{c}) +  p_\text{ext}N_\text{ext}k_\text{ext}  + p_\text{c} k_\text{c},
\end{align}
where $N_\text{ext}$ is the number of external rings, $k_\text{ext}$ is the transfer rate constant between one external B850 and the initially excited B850, and $k_\text{c}$ is the rate constant for energy transfer between the polaritons and the initially excited B850. For simplicity, the population of a single external B850, $p_\text{ext}$, is assumed to be the same for each of the external B850s.

The equation for the polaritons takes the form
\begin{align}
   \frac{d}{dt} p_\text{c} = -p_c(k_\text{c} + k_\text{c}N_\text{ext}) + k_\text{c}(p_\text{B850} + p_\text{ext}N_\text{ext}).
\end{align}

Because the overlap sum between the external B850s and the polaritons is very large (Figure \ref{mainoverlaps}), a steady state for the polariton population is quickly reached due to the equilibration between the populations of external B850s and the polaritons. Hence, $\frac{d}{dt} p_\text{c} \approx 0$, which allows solving for $p_\text{c}$ as 
\begin{align}\label{cavityeq}
  p_\text{c}=  \frac{p_\text{B850}+p_\text{ext}N_\text{ext}}{1+ N_\text{ex}}.
\end{align}

Inserting Equation \ref{cavityeq} into Equation \ref{b850transfereq} and using $p_\text{ext} N_\text{ext}   = 1 - p_\text{c}- p_\text{B850}$ yields
\begin{align}
\frac{d}{dt} p_\text{B850} = - p_\text{B850} (  k_\text{ext} (N_\text{ext} +1) + k_\text{c}) - p_\text{c}( k_\text{ext}  + \frac{k_\text{c}  }{1+ N_\text{ex}}) +  k_\text{ext}  + \frac{k_\text{c}  }{1+ N_\text{ext}}. 
\end{align}

Taking a derivative and using $d p_\text{c} /dt=0$ leads to
\begin{align}
\frac{d^2}{dt^2} p_\text{B850} + \frac{d}{dt} p_\text{B850} (  k_\text{ext} (N_\text{ext} +1) + k_\text{c}) =0. 
\end{align}
The characteristic equation for the solution of the differential equation gives the roots 0 and $k_\text{ext}N_\text{LH2} - k_\text{c}$, where $N_\text{LH2} = N_\text{ext} + 1$. With the conditions $p_\text{B850}(0) = 1$ and $p_\text{B850}(\infty) = 1/N_\text{LH2}$, the solution takes the form
\begin{equation}
    p_\text{B850}(t) = \frac{1+e^{-(k_\text{ext}N_\text{LH2} + k_c)t }(N_\text{LH2}-1)}{N_\text{LH2}}.
\end{equation}
The rate for the energy transfer in the limit of multiple LH2s can thus be estimated as
\begin{equation}
    k = k_\text{ext}N_\text{LH2} + k_\text{c}.
\end{equation}

\end{document}